\documentclass[
    reprint,
    showpacs,preprintnumbers,amsmath,amssymb,superscriptaddress,
    floatfix, 
    nobalancelastpage,
    aps,prc
    ]
{revtex4-1}
\usepackage{graphicx}
\usepackage{pdfcomment}
\usepackage{dcolumn}
\usepackage{bm}
\usepackage{float}
\usepackage{multirow, booktabs}
\usepackage{lineno}
\usepackage{bibentry}
\usepackage{color}
\usepackage{epstopdf,epsfig}
\usepackage{soul}
\def\apjs{{ApJS}} 

\begin{document}

\preprint{APS/123}

\title{Low-lying level structure of $^{56}$Cu and its implications on the \mbox{rp process}}
  
\author{W-J.~Ong} 
\affiliation{National Superconducting Cyclotron Laboratory, Michigan State University, East Lansing, MI 48824, USA}
\affiliation{Joint Institute for Nuclear Astrophysics, Michigan State University, East Lansing, MI 48824, USA}
\affiliation{Department of Physics and Astronomy, Michigan State University, East Lansing, MI 48824, USA}  
\author{C.~Langer}
\affiliation{National Superconducting Cyclotron Laboratory, Michigan State University, East Lansing, MI 48824, USA}
\affiliation{Joint Institute for Nuclear Astrophysics, Michigan State University, East Lansing, MI 48824, USA}
\author{F.~Montes}
\affiliation{National Superconducting Cyclotron Laboratory, Michigan State University, East Lansing, MI 48824, USA}
\affiliation{Joint Institute for Nuclear Astrophysics, Michigan State University, East Lansing, MI 48824, USA}
\author{A.~Aprahamian}
\affiliation{Department of Physics and Joint Institute for Nuclear Astrophysics, University of Notre Dame, Notre Dame, Indiana 46556, USA}
\author{D.~W.~Bardayan}
\altaffiliation[Present address: ]{Department of Physics, University of Notre Dame, Notre Dame, Indiana 46556, USA}
\affiliation{Physics Division, Oak Ridge National Laboratory, Oak Ridge, Tennessee 37831, USA}
\author{D.~Bazin}
\affiliation{National Superconducting Cyclotron Laboratory, Michigan State University, East Lansing, MI 48824, USA}
\author{B.~A. Brown}
\affiliation{National Superconducting Cyclotron Laboratory, Michigan State University, East Lansing, MI 48824, USA}
\affiliation{Department of Physics and Astronomy, Michigan State University, East Lansing, MI 48824, USA}
\author{J.~Browne}
\affiliation{National Superconducting Cyclotron Laboratory, Michigan State University, East Lansing, MI 48824, USA}
\affiliation{Joint Institute for Nuclear Astrophysics, Michigan State University, East Lansing, MI 48824, USA}
\affiliation{Department of Physics and Astronomy, Michigan State University, East Lansing, MI 48824, USA}
\author{H.~Crawford}
\affiliation{Nuclear Science Division, Lawrence Berkeley National Laboratory, Berkeley CA, 94720, USA}
\author{R.~Cyburt}
\affiliation{National Superconducting Cyclotron Laboratory, Michigan State University, East Lansing, MI 48824, USA}
\affiliation{Joint Institute for Nuclear Astrophysics, Michigan State University, East Lansing, MI 48824, USA}
\author{E.~B.~Deleeuw}
\affiliation{National Superconducting Cyclotron Laboratory, Michigan State University, East Lansing, MI 48824, USA}
\affiliation{Joint Institute for Nuclear Astrophysics, Michigan State University, East Lansing, MI 48824, USA}
\affiliation{Department of Physics and Astronomy, Michigan State University, East Lansing, MI 48824, USA}  
\author{C.~Domingo-Pardo}
\affiliation{IFIC, CSIC-University of Valencia, E-46071 Valencia, Spain}
\author{A.~Gade}
\affiliation{National Superconducting Cyclotron Laboratory, Michigan State University, East Lansing, MI 48824, USA}
\affiliation{Department of Physics and Astronomy, Michigan State University, East Lansing, MI 48824, USA}
\author{S.~George}
\altaffiliation{Max-Planck-Institut f\"ur Kernphysik, 69117 Heidelberg, Germany}
\affiliation{Institut f. Physik, Ernst-Moritz-Arndt-Universit\"at, 17487 Greifswald, Germany}
\author{P.~Hosmer}
\affiliation{Department of Physics, Hillsdale College, Hillsdale, MI 49242, USA}
\author{L.~Keek}
\affiliation{National Superconducting Cyclotron Laboratory, Michigan State University, East Lansing, MI 48824, USA}
\affiliation{Joint Institute for Nuclear Astrophysics, Michigan State University, East Lansing, MI 48824, USA}
\affiliation{Department of Physics and Astronomy, Michigan State University, East Lansing, MI 48824, USA}
\author{A.~Kontos}
\affiliation{National Superconducting Cyclotron Laboratory, Michigan State University, East Lansing, MI 48824, USA}
\affiliation{Joint Institute for Nuclear Astrophysics, Michigan State University, East Lansing, MI 48824, USA}
\author{I-Y.~Lee}
\affiliation{Nuclear Science Division, Lawrence Berkeley National Laboratory, Berkeley CA, 94720, USA}
\author{A.~Lemasson}
\affiliation{National Superconducting Cyclotron Laboratory, Michigan State University, East Lansing, MI 48824, USA}
\author{E.~Lunderberg}
\affiliation{National Superconducting Cyclotron Laboratory, Michigan State University, East Lansing, MI 48824, USA}
\affiliation{Department of Physics and Astronomy, Michigan State University, East Lansing, MI 48824, USA}
\author{Y.~Maeda}
\affiliation{Department of Applied Physics, University of Miyazaki, Miyazaki, Miyazaki 889-2192, Japan}
\author{M.~Matos}
\affiliation{Department of Physics and Astronomy, Louisiana State University, Baton Rouge, Louisiana 70803-4001, USA}
\author{Z.~Meisel}
\affiliation{National Superconducting Cyclotron Laboratory, Michigan State University, East Lansing, MI 48824, USA}
\affiliation{Joint Institute for Nuclear Astrophysics, Michigan State University, East Lansing, MI 48824, USA}
\affiliation{Department of Physics and Astronomy, Michigan State University, East Lansing, MI 48824, USA}
\author{S.~Noji}
\affiliation{National Superconducting Cyclotron Laboratory, Michigan State University, East Lansing, MI 48824, USA}
\author{F.~M.~Nunes}
\affiliation{National Superconducting Cyclotron Laboratory, Michigan State University, East Lansing, MI 48824, USA}
\affiliation{Department of Physics and Astronomy, Michigan State University, East Lansing, MI 48824, USA}
\author{A.~Nystrom}
\affiliation{Department of Physics and Joint Institute for Nuclear Astrophysics, University of Notre Dame, Notre Dame, Indiana 46556, USA}
\author{G.~Perdikakis}
\affiliation{Department of Physics, Central Michigan University, Mt. Pleasant, MI 48859, USA}
\affiliation{National Superconducting Cyclotron Laboratory, Michigan State University, East Lansing, MI 48824, USA}
\affiliation{Joint Institute for Nuclear Astrophysics, Michigan State University, East Lansing, MI 48824, USA}
\author{J.~Pereira}
\affiliation{National Superconducting Cyclotron Laboratory, Michigan State University, East Lansing, MI 48824, USA}
\affiliation{Joint Institute for Nuclear Astrophysics, Michigan State University, East Lansing, MI 48824, USA}
\author{S.~J.~Quinn}
\affiliation{National Superconducting Cyclotron Laboratory, Michigan State University, East Lansing, MI 48824, USA}
\affiliation{Joint Institute for Nuclear Astrophysics, Michigan State University, East Lansing, MI 48824, USA}
\affiliation{Department of Physics and Astronomy, Michigan State University, East Lansing, MI 48824, USA}
\author{F.~Recchia}
\affiliation{National Superconducting Cyclotron Laboratory, Michigan State University, East Lansing, MI 48824, USA}
\author{H.~Schatz}
\affiliation{National Superconducting Cyclotron Laboratory, Michigan State University, East Lansing, MI 48824, USA}
\affiliation{Joint Institute for Nuclear Astrophysics, Michigan State University, East Lansing, MI 48824, USA}
\affiliation{Department of Physics and Astronomy, Michigan State University, East Lansing, MI 48824, USA}
\author{M.~Scott}
\affiliation{National Superconducting Cyclotron Laboratory, Michigan State University, East Lansing, MI 48824, USA}
\affiliation{Joint Institute for Nuclear Astrophysics, Michigan State University, East Lansing, MI 48824, USA}
\affiliation{Department of Physics and Astronomy, Michigan State University, East Lansing, MI 48824, USA}
\author{K.~Siegl}
\affiliation{Department of Physics and Joint Institute for Nuclear Astrophysics, University of Notre Dame, Notre Dame, Indiana 46556, USA}
\author{A.~Simon}
\altaffiliation{Gottwald Center for the Sciences, University of Richmond, 28 Westhampton Way, Richmond, VA 23173}
\affiliation{National Superconducting Cyclotron Laboratory, Michigan State University, East Lansing, MI 48824, USA}
\affiliation{Joint Institute for Nuclear Astrophysics, Michigan State University, East Lansing, MI 48824, USA}
\author{M.~Smith}
\affiliation{Department of Physics and Joint Institute for Nuclear Astrophysics, University of Notre Dame, Notre Dame, Indiana 46556, USA}
\author{A.~Spyrou}
\affiliation{National Superconducting Cyclotron Laboratory, Michigan State University, East Lansing, MI 48824, USA}
\affiliation{Joint Institute for Nuclear Astrophysics, Michigan State University, East Lansing, MI 48824, USA}
\affiliation{Department of Physics and Astronomy, Michigan State University, East Lansing, MI 48824, USA}
\author{J.~Stevens}
\affiliation{National Superconducting Cyclotron Laboratory, Michigan State University, East Lansing, MI 48824, USA}
\affiliation{Joint Institute for Nuclear Astrophysics, Michigan State University, East Lansing, MI 48824, USA}
\affiliation{Department of Physics and Astronomy, Michigan State University, East Lansing, MI 48824, USA}
\author{S.~R.~Stroberg}
\affiliation{National Superconducting Cyclotron Laboratory, Michigan State University, East Lansing, MI 48824, USA}
\affiliation{Department of Physics and Astronomy, Michigan State University, East Lansing, MI 48824, USA}
\author{D.~Weisshaar}
\affiliation{National Superconducting Cyclotron Laboratory, Michigan State University, East Lansing, MI 48824, USA}
\author{J.~Wheeler}
\affiliation{National Superconducting Cyclotron Laboratory, Michigan State University, East Lansing, MI 48824, USA}
\affiliation{Joint Institute for Nuclear Astrophysics, Michigan State University, East Lansing, MI 48824, USA}
\affiliation{Department of Physics and Astronomy, Michigan State University, East Lansing, MI 48824, USA}
\author{K.~Wimmer}
\affiliation{Department of Physics, Central Michigan University, Mt. Pleasant, MI 48859, USA}
\affiliation{National Superconducting Cyclotron Laboratory, Michigan State University, East Lansing, MI 48824, USA}
\author{R.~G.~T.~Zegers}
\affiliation{National Superconducting Cyclotron Laboratory, Michigan State University, East Lansing, MI 48824, USA}
\affiliation{Joint Institute for Nuclear Astrophysics, Michigan State University, East Lansing, MI 48824, USA}
\affiliation{Department of Physics and Astronomy, Michigan State University, East Lansing, MI 48824, USA}
\date{\today}  
\begin{abstract}
The low-lying energy levels of proton-rich $^{56}$Cu have been extracted using in-beam $\gamma$-ray spectroscopy with the state-of-the-art $\gamma$-ray tracking array GRETINA in conjunction with the S800 spectrograph at the National Superconducting Cyclotron Laboratory at Michigan State University. Excited states in $^{56}$Cu serve as resonances in the $^{55}$Ni(p,$\gamma$)$^{56}$Cu reaction, which is a part of the rp-process in type I x-ray bursts. To resolve existing ambiguities in the reaction Q-value, a more localized IMME mass fit is used resulting in $Q=639\pm82$~keV.
We derive the first experimentally-constrained thermonuclear reaction rate for $^{55}$Ni(p,$\gamma$)$^{56}$Cu. We find that, with this new rate, the rp-process may bypass the $^{56}$Ni waiting point via the $^{55}$Ni(p,$\gamma$) reaction for typical x-ray burst conditions with a branching of up to $\sim$40$\%$. We also identify additional nuclear physics uncertainties that need to be addressed before drawing final conclusions about the rp-process reaction flow in the $^{56}$Ni region. 
\end{abstract} 
                                                                                                                                                                                      
\maketitle 

\section{Introduction}
Accreting neutron stars in binary systems undergo episodes of explosive hydrogen and helium burning, observed as Type-I x-ray bursts.  The main observable of these events, the x-ray burst light-curve, is shaped by the nuclear energy generation during the rapid proton-capture process (rp-process) \cite{1981ApJS...45..389W, 1998PhR...294..167S}. This process involves a series of proton captures and $\beta$-decays that proceed near the proton-drip line. 

\begin{figure}[h]
\centering
\includegraphics[trim= 5cm 4cm 8cm 2cm, clip=True, scale=0.40]{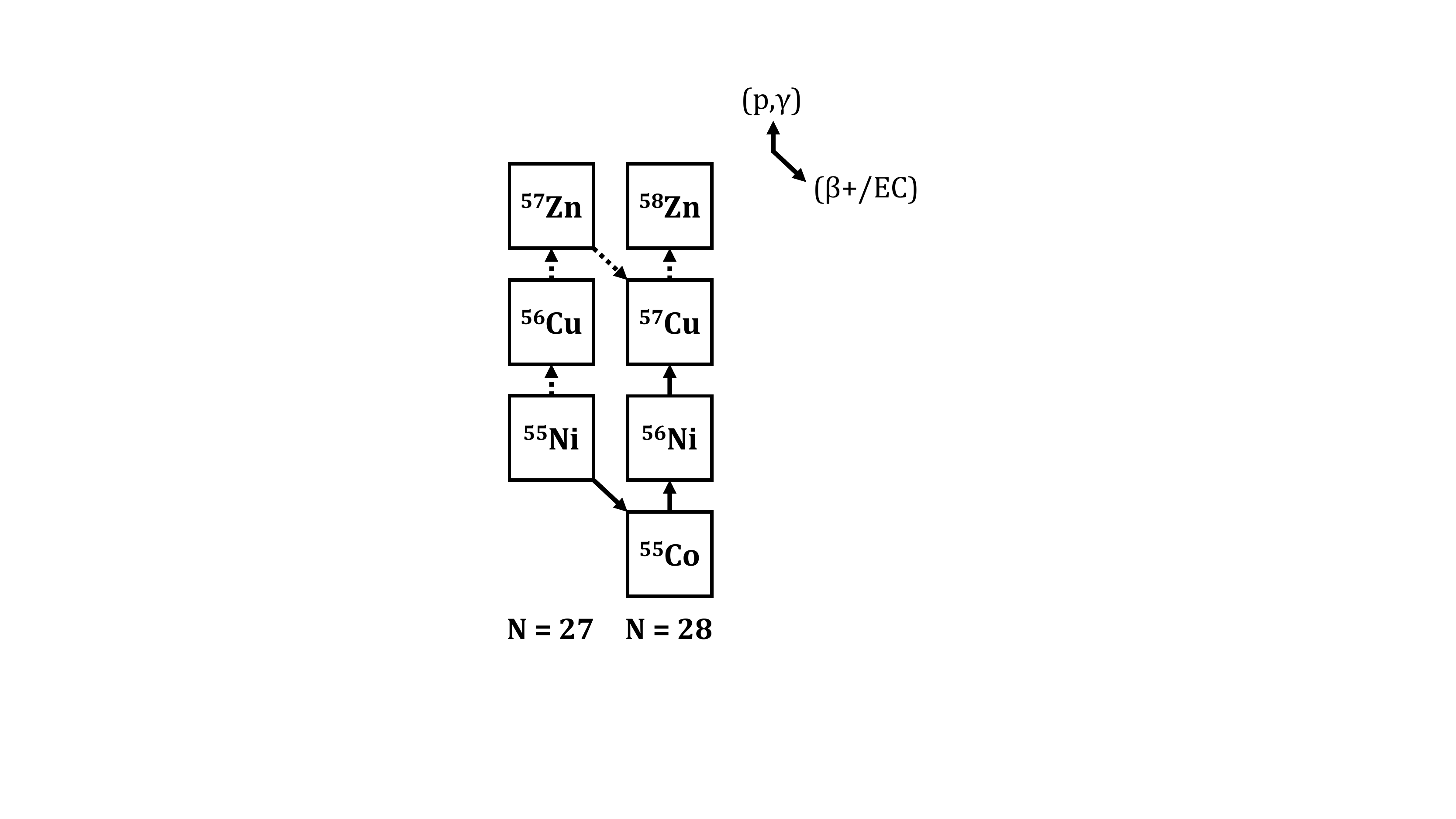}
\caption{The nuclide chart in the region of the $^{56}$Ni waiting point. The conventional $rp$-process flow leading to $^{56}$Ni is denoted by the solid line. The potential bypass, sequential proton-captures along the N~=~27 isotonic chain, is denoted by the dashed line.}
\label{bypassflow}
\end{figure}

Reaction rates connected to the so-called waiting point nuclei \cite{1994ApJ...432..326V}, where the reaction flow slows down significantly, have the most significant impact on the observed light curve. The doubly-magic nucleus $^{56}$Ni has been identified as one of a few major waiting points in the rp-process. This is due to the combination of its long stellar electron capture half-life of $\sim$3~hrs \cite{1982ApJ...252..715F} (which for the fully ionized ion differs from its terrestrial half-life and depends on the stellar electron density), and its low proton-capture Q-value (690~keV) \cite{doi:1880784}. The effective lifetime of $^{56}$Ni under typical x-ray burst conditions, which depends steeply on temperature, has been constrained by experimental data related to the $^{56}$Ni(p,$\gamma$) \cite{1998PhRvL..80..676R, 2001PhRvC..64d5801F}  and $^{57}$Cu(p,$\gamma$) \cite{PhysRevLett.113.032502} reaction rates. However, large uncertainties exist in the nuclear physics of more neutron-deficient nuclei in the $^{56}$Ni region. In particular, a sequence of proton-capture reactions in the $^{55}$Ni, $^{56}$Cu, $^{57}$Zn isotonic chain may be strong enough for the rp-process to bypass $^{56}$Ni (Fig. \ref{bypassflow}). In this case, $^{56}$Ni would not be an rp-process waiting point, reducing the sensitivity of burst models to the $^{56}$Ni(p,$\gamma$) rate. The $^{57}$Cu(p,$\gamma$) reaction rate remains important because the bypass exits the N=27 isotonic chain through $\beta$-decay of $^{57}$Zn to $^{57}$Cu. Consequently, the reaction flow would proceed more rapidly into the Ge-Se-Kr mass region and a lower amount of A~=~56 material would be produced in the ashes. 
 
The $^{55}$Ni(p,$\gamma$) reaction determines the branching at $^{55}$Ni into the $^{56}$Ni bypass reaction sequence. Here, we address uncertainties in this reaction rate experimentally, and reanalyze theoretical predictions of the reaction Q-value. We then use the new data to determine, in the context of the remaining nuclear physics uncertainties, the conditions under which the rp-process bypasses $^{56}$Ni.    

The $^{55}$Ni(p,$\gamma$)$^{56}$Cu reaction proceeds through a few isolated narrow resonances, and the astrophysical rate can be approximated by
\begin{equation}
N_A \langle \sigma \upsilon \rangle \propto \sum_i (\omega \gamma)_i \mathrm{exp}(-E_i/kT)
\label{rateeq}
\end{equation}
where $E_i = E_{i}^{x} - Q$ is the resonance energy with reaction Q-value $Q$ and $^{56}$Cu exitation energy $E_i^x$. The resonance strength is given by
\begin{equation}
\omega \gamma = \frac{2J + 1}{(2J_p + 1)(2J_{^{55}\mathrm{ Ni}} + 1)} \frac{\Gamma_p \Gamma_\gamma}{\Gamma}.
\end{equation}
Here, $J$ is the resonance spin, $J_p$ the proton spin, $J_{^{55}\mathrm{Ni}}$ is the ground-sate spin of $^{55}$Ni, $\Gamma_p$ the proton partial width, $\Gamma_\gamma$ the $\gamma$ partial width and $\Gamma=\Gamma_p+\Gamma_\gamma$. 

Only scarce experimental data for the odd-odd $^{56}$Cu nucleus exist in the literature. $^{56}$Cu, as well as its well-understood mirror nucleus $^{56}$Co, are part of the $A=56$, $T=1$ isospin triplet. Based on this, the ground-state of $^{56}$Cu is assumed to be $J^\pi=4^+$ with a measured terrestrial $\beta$-decay half-life of 93(3)~ms \cite{Junde20111513}. To date, no low-lying excited states have been observed experimentally. In a recent $\beta$-delayed proton decay study of $^{56}$Zn, several higher-lying $^{56}$Cu resonances above 1391~keV excitation energy were observed \cite{Orrigo}. Under astrophysical conditions, however, these resonances are too high in energy to be of relevance.  In the absence of knowledge of spectroscopic information, shell-model calculations using the KB3 interaction in the $pf$-shell performed with the code ANTOINE have been used in the past \cite{Fisker2001453}. However, uncertainties in shell-model predictions of excitation energies can amount up to 200~keV, leading to orders of magnitude uncertainty in the resonant-capture rate. Here we experimentally determine, for the first time, the excitation energies of low-lying states in $^{56}$Cu that serve as resonances in the $^{55}$Ni(p,$\gamma$)$^{56}$Cu reaction

For a precise determination of the $^{55}$Ni(p,$\gamma$)$^{56}$Cu rate, both the low-lying level scheme of $^{56}$Cu and the reaction Q-value need to be well-known since the resonance energies enter the rate exponentially. While the mass of $^{55}$Ni is experimentally well-known with an error of 0.75~keV \cite{PhysRevC.82.034311}, the mass of $^{56}$Cu is not experimentally known. Conflicting predictions for the $^{56}$Cu mass exist in the literature. The extrapolated $^{56}$Cu mass in the AME2003 compilation \cite{Wapstra2003129} results in a $^{55}$Ni proton-capture Q-value of 560(140)~keV. A similar result of 600(100)~keV is obtained with Coulomb shift calculations \cite{PhysRevC.65.045802}. Using the $^{56}$Cu mass in the most recent AME2012 compilation, however, results in a Q-value of 190(200)~keV \cite{doi:1880784}. We obtain a new prediction for the reaction Q-value by using the \textit{isobaric multiplet mass equation} (IMME). 

\section{Experimental Determination of the $^{56}$Cu level scheme}
\begin{figure}[h]
\centering

\includegraphics[trim = 0.05cm 0.05cm 0.05cm 0.05cm, clip=True, width=0.50\textwidth] {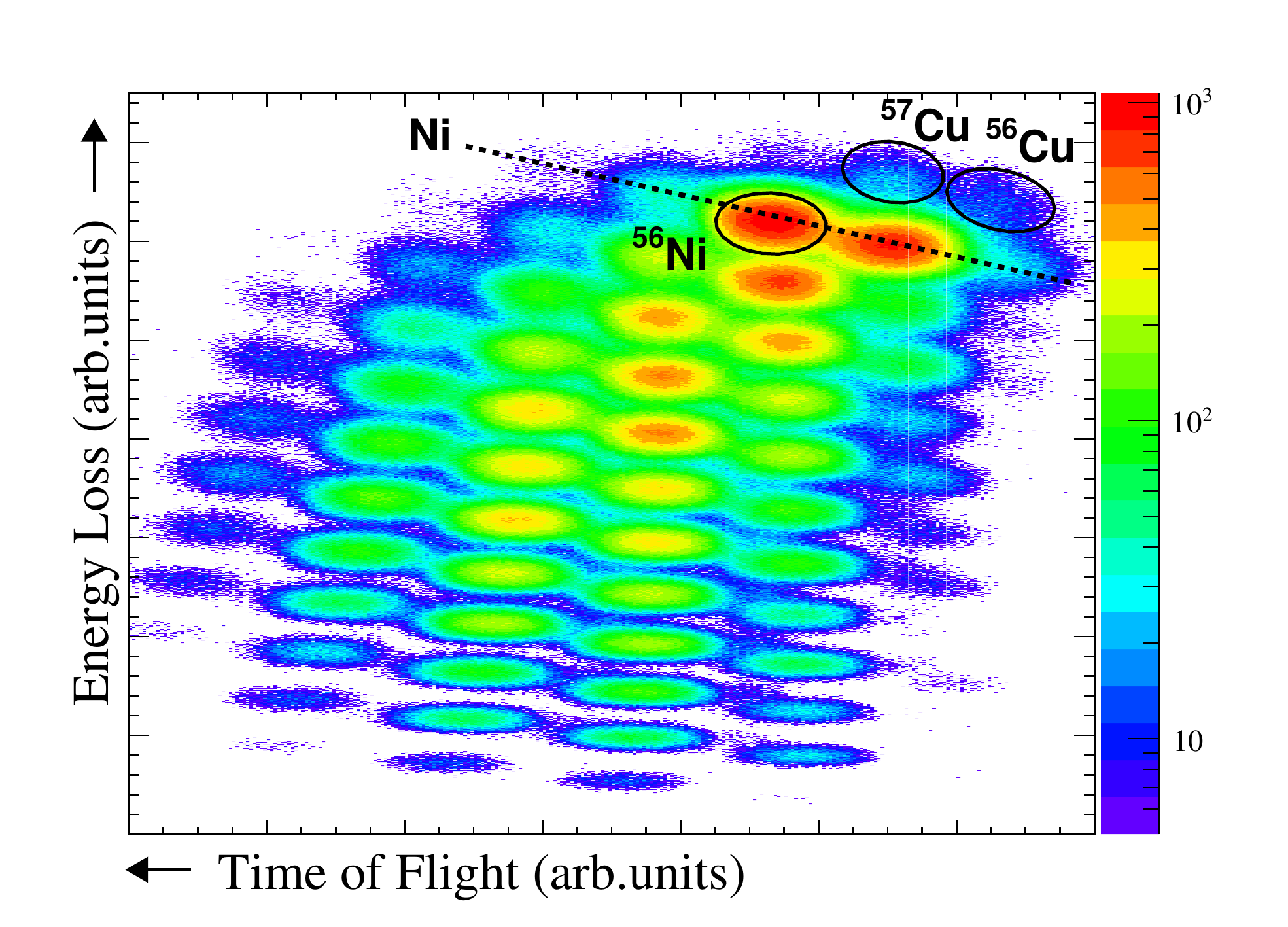}
\caption{(Color online:) $\Delta E$-TOF particle identification for ions reaching the S800 focal plane. Color indicates the number of counts per bin. The Ni isotopic chain (dotted line) and the $^{56}$Ni (leftmost ellipse), $^{56}$Cu (rightmost ellipse), and $^{57}$Cu (middle ellipse) isotopes are also marked (not actual analysis gates).}
\label{PID}
\end{figure}

Excited states of $^{56}$Cu were populated in inverse kinematics in an experiment performed at the National Superconducting Cyclotron Laboratory (NSCL) at Michigan State University \cite{PhysRevLett.113.032502}. A stable 160~MeV/$u$ $^{58}$Ni primary beam impinged on a 752~mg/cm$^2$ $^{9}$Be target placed at the entrance of the A1900 fragment separator \cite{Morrissey200390}. After purification by the A1900 using the $B\rho-\Delta E-B\rho$ method, the produced $^{56}$Ni secondary beam had a rate of $\sim10^5$~pps with a beam purity of $\sim75\%$. The $^{56}$Ni beam (E~=~$\sim75$~MeV/u) was then incident upon a 225 mg/cm$^2$ CD$_2$ target, producing $^{56}$Cu through various reaction channels. The CD$_2$ target was located in the center of the $\gamma$-ray energy tracking array GRETINA \cite{2013NIMPA.709...44P}, which was used to measure energies of the prompt $\gamma$-rays emitted from the de-excitation of the excited states in $^{56}$Cu. GRETINA consists of 28 coaxial HPGe detector crystals, which are closely-packed to cover roughly 1$\pi$ in solid angle. Kinematical reconstruction of the momentum, angle, and position of each $^{56}$Cu recoil at the target based on observables at the S800 focal plane, combined with the high position resolution for $\gamma$-ray detection in GRETINA allow for accurate Doppler-shift corrections for $\gamma$-rays emitted in-flight. The recoil velocity $\beta = \upsilon/c$ used for the Doppler-shift correction was extracted using momentum information, and was determined for each individual event to correct for energy loss in the target. The $^{56}$Cu recoils, after leaving the target, were identified using detectors situated in the focal plane of the S800 spectrograph \cite{Bazin2003629} located downstream from GRETINA. The S800 focal plane contained  a set of two cathode readout drift counters that were used to determine the particle trajectory, a gas-filled ionization chamber that measured energy loss $\Delta E$, and a plastic scintillator that, along with the thin timing scintillator at the A1900 focal plane and the scintillator at the S800 object position, were used for time-of-flight (TOF) analysis. The measured time-of-flight between the A1900 focal plane and S800 object scintillators was used to uniquely identify the $^{56}$Cu recoil by $\Delta E$-TOF (Fig. \ref{PID}). 

\begin{figure}[H]
\includegraphics[width=0.47\textwidth]{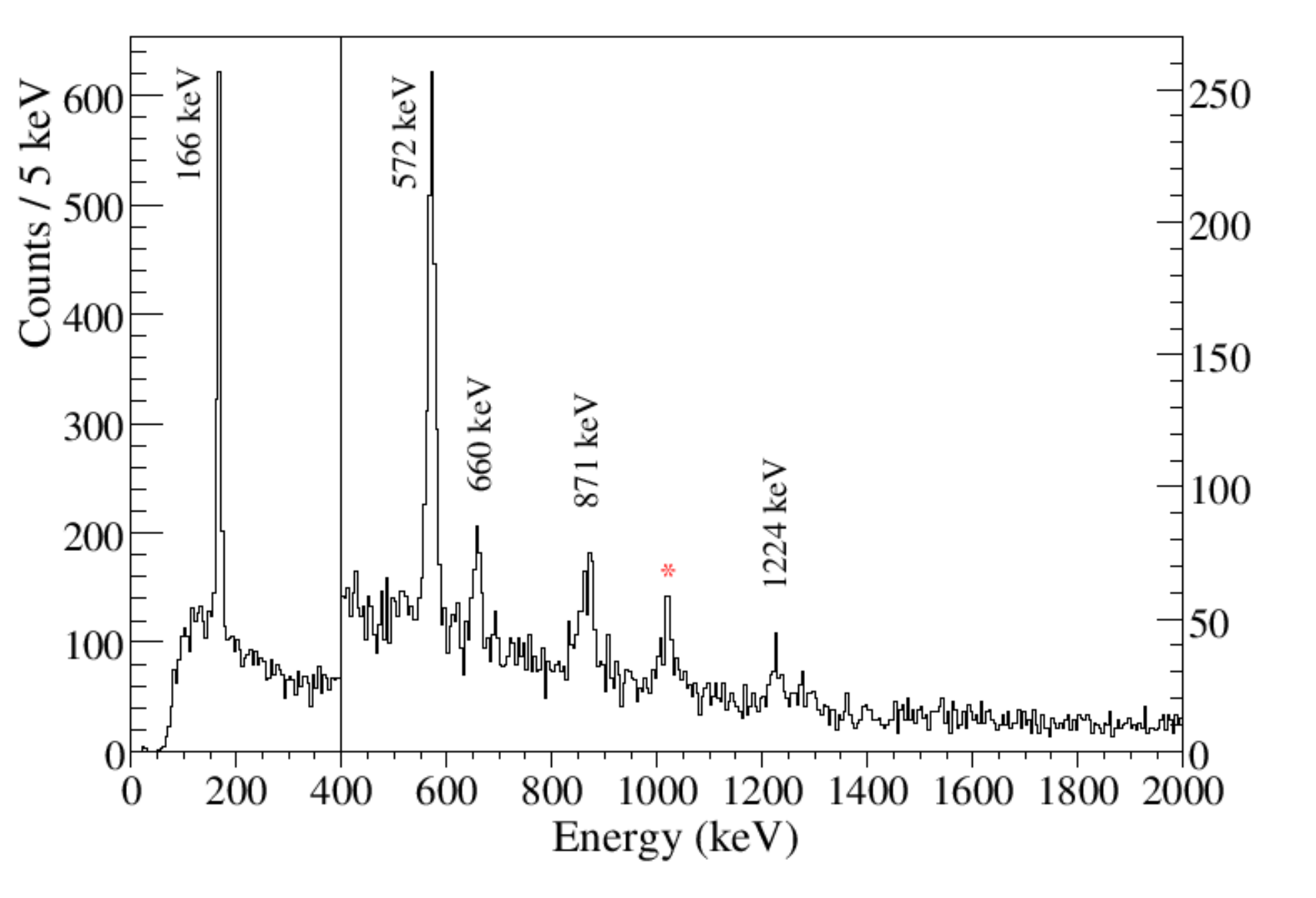}
\caption{Doppler-corrected $\gamma$-ray spectrum measured with GRETINA in coincidence with $^{56}$Cu ions in the S800 focal plane. A nearest neighbor addback algorithm has been applied. The asterisk indicates contamination from $^{57}$Cu.}
\label{gammas} 
\end{figure}

The low-lying level scheme of $^{56}$Cu was constructed using observed $\gamma$-ray transitions, $\gamma-\gamma$ coincidences and guidance from the experimentally based level scheme of the mirror nucleus $^{56}$Co.  The Doppler-corrected spectrum of the $\gamma$-rays detected by GRETINA, in coincidence with the $^{56}$Cu recoils in the S800, shows five $\gamma$-ray transitions (Fig.~\ref{gammas}). An additional line at $E_\gamma$~=~1027~keV stems from contamination from a well-known $\gamma$-ray transition in $^{57}$Cu which is located next to $^{56}$Cu in the particle identification spectrum (Fig. \ref{PID}). We confirmed that this $\gamma$-ray line disappears from the $\gamma$ spectrum when the particle identification gate in Fig.~\ref{PID} is tightened to only include the most centrally located events in the $^{56}$Cu  recoil region.

\begin{figure}[h]
\includegraphics[trim=0cm 0cm 1cm 0cm, width=0.5\textwidth]{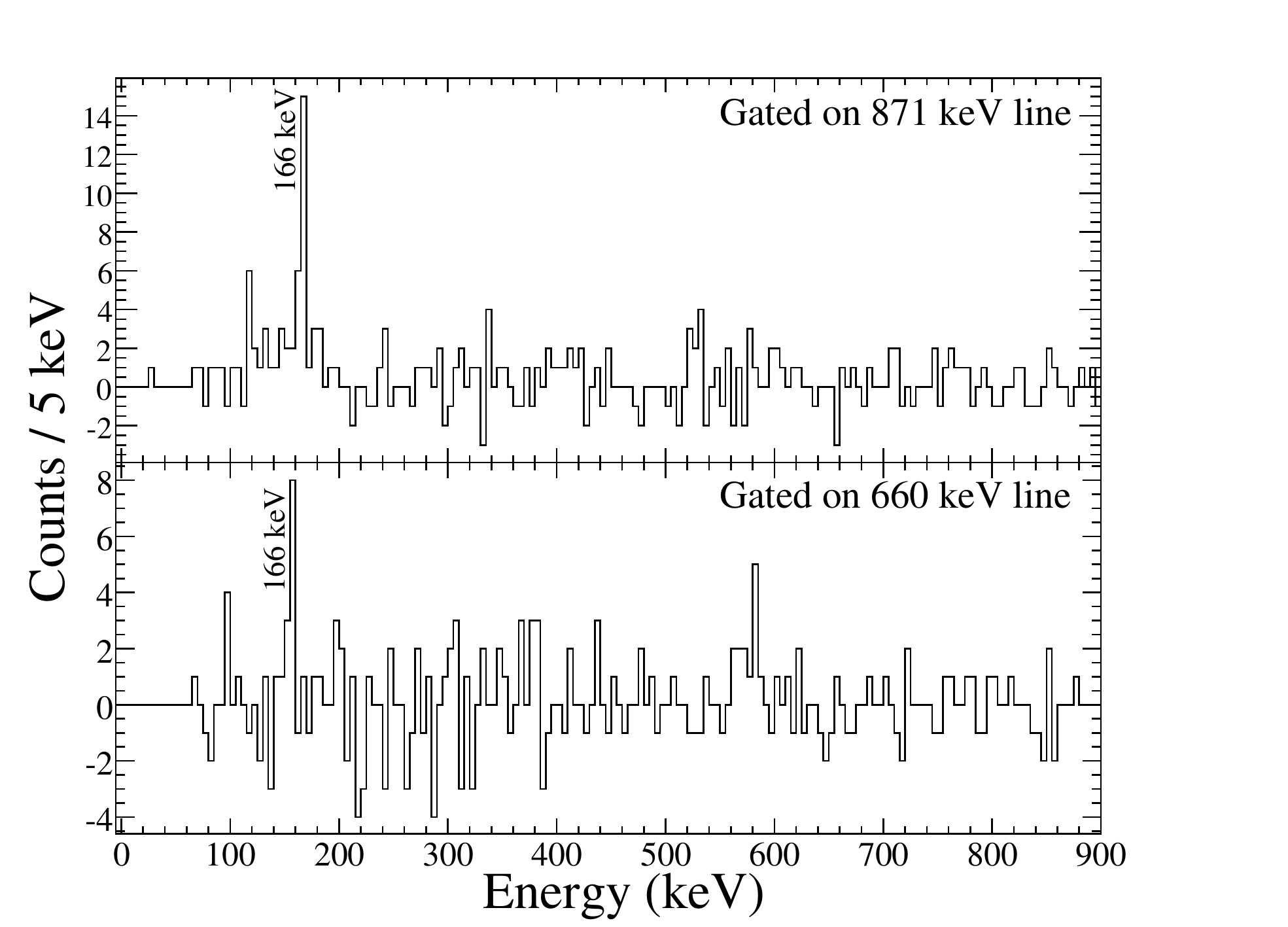}
\caption{$\gamma-\gamma$ coincidences with $E_\gamma$ = 871 (3)~keV (upper panel) and $E_\gamma$ = 660 (3)~keV (lower panel).}
\label{coincidence}
\end{figure}

\begin{figure}[h]
\includegraphics[trim=4cm 4.5cm 1cm 5.5cm,clip, width=0.5\textwidth]{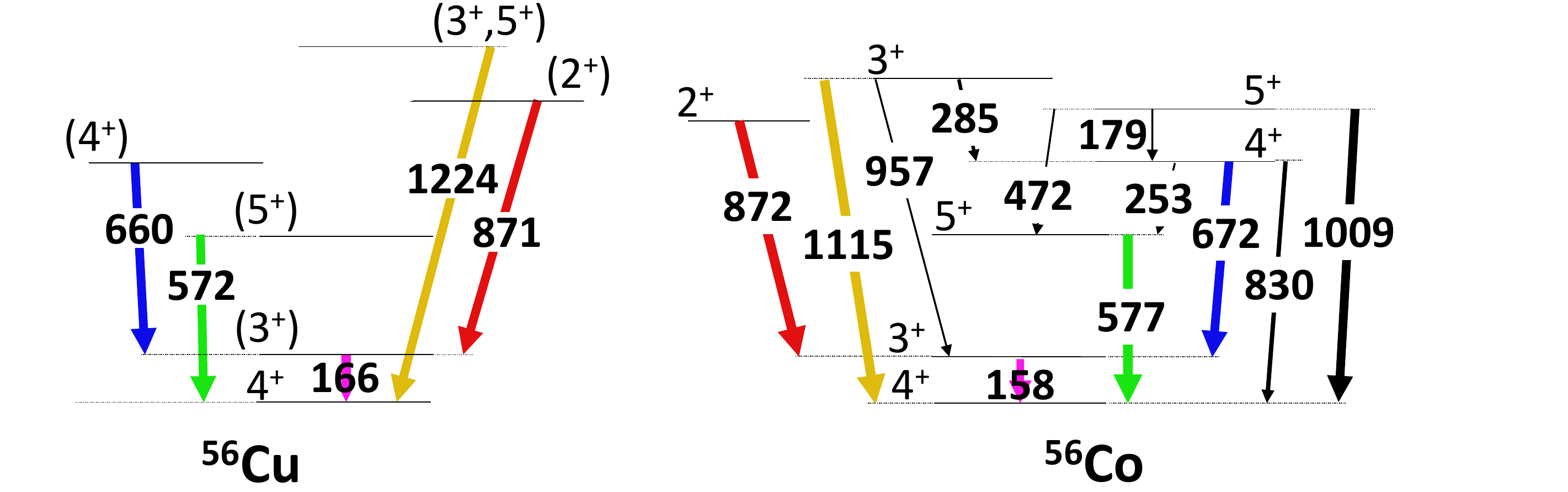}
\caption{(Color online:) Proposed low-lying level scheme of $^{56}$Cu (left) in comparison to its mirror nucleus $^{56}$Co (right). Tentative spin and parity assignments are shown in parentheses. The observed $\gamma$-transitions are shown, with the corresponding transitions in the mirror shown with the same color.}
\label{gf3level}
\end{figure}

The left half of figure \ref{gf3level} shows the reconstructed $^{56}$Cu level scheme. The strongest observed line at $E_\gamma$~=~166(1)~keV is close in energy to the first excited state at 158~keV ($J^\pi = 3^+$) in the mirror nucleus $^{56}$Co. Based on experimental information from the $^{56}$Co mirror nucleus, we expect the first excited state to be the most intense transition as it is fed from several higher-lying states. This line is observed to be in coincidence with two other $\gamma$-transitions, supporting its assignment as direct decay from the the first excited $3^+$ state (Fig. \ref{coincidence}). 

The transitions at $E_\gamma$~=~660(3)~keV and $E_\gamma$~=~871(3)~keV are observed to be in coincidence with the $E_\gamma$~=~166(1)~keV transition as shown in Fig. \ref{coincidence}, but not with each other. Based on the prior assignment of the 166~keV first excited state, two states are placed at $E_x=826(3)$~keV and $E_x=1037(3)$~keV, respectively.  No ground state decays are observed for either of these states.  There are three known states in the $^{56}$Co mirror at similar energies of $E_x=830, 970$ and $1009$~keV. Of those, the 1009~keV state decays predominantly to the ground state.  Both the $J^{\pi}=4^+$ 830~keV and $J^\pi=2^+$ 970~keV states decay primarily to the first excited state at 158~keV with only a 34$\%$ and 0.3$\%$ direct transition to the ground state, respectively.  Based on the decay modes and similarities in energies, the two observed states at $E_x=826(3)$~keV and $E_x=1037(3)$~keV are tentatively assigned as $J^{\pi}=4^+$ and $J^\pi=2^+$, respectively.

The observed line at $E_\gamma$~=~572(1)~keV is not seen in coincidence with the 166~keV line. The mirror $^{56}$Co has a $J^\pi~=~5^+$ state at  $E_x~=~577$~keV that decays only to the ground state. Based on the similar energies and similar decay modes, we tentatively assign the 572~keV transition to be the second excited $J^\pi~=~5^+$ state.

The $E_\gamma=1224(4)$~keV line is not observed to be in coincidence with any other $\gamma$-ray transition, and it is therefore assigned to a level at that energy. The analog states in the mirror with the closest energies are 1009~keV ($J^\pi=5_2^+$) and 1115~keV ($J^\pi=3_2^+$) which both decay largely to the ground state. Other higher lying states in $^{56}$Co (the next one is at 1450~keV) decay predominantly through cascades, which is not supported by our measurement. We tentatively assign $E_x~=~1224(4)$~keV as either the $J^\pi$~=~3$_2^+$ or the $J^\pi=5_2^+$ state. The observed transitions, intensities and assignments are tabulated in Table~\ref{tablegam}. A comparison to the mirror nucleus is shown in Fig.~\ref{gf3level}. 


\begin{table}[h]
\caption{Reconstructed level scheme of  $^{56}$Cu excitation levels with observed transition energies ($E_\gamma$), relative intensities ($I_\gamma$) normalized to the $E_\gamma$~=~166 keV line, and tentative spin-parity assignments (see text for details).}
\centering
\begin{ruledtabular}
\begin{tabular}{c c c c}
$E_x$ (keV) & $E_{\gamma}$ (keV) & $I_{\gamma}$ ($\%$) & $J_i^\pi \rightarrow J_f^\pi$ \\ \hline
166 (1) & 166 (1) & 100 & $(3_1^+) \rightarrow \text{g.s.}$ \\
572 (1) & 572 (1) & 122 (8) & $(5_1^+) \rightarrow \text{g.s.}$  \\
826 (3) & 660 (3) & 28 (8) & $(4_2^+) \rightarrow (3^+_1)$  \\
1037 (3) & 871 (3) & 50 (8) & $(2_1^+) \rightarrow (3^+_1)$ \\
1224 (4) & 1224 (4) & 19 (10) & $(3_2^+,5_2^+) \rightarrow \text{g.s.}$
\label{tablegam}
\end{tabular}
\end{ruledtabular}
\end{table}

\section{Mass Estimate of $^{56}$Cu using the Isobaric Multiplet Mass Equation}
We use the isobaric mass multiplet equation (IMME) to predict a new $^{56}$Cu mass, which is needed to derive the reaction Q-value and the resonance energies. The $^{56}$Cu ground state (J$^{\pi}=4^+$) is part of the $A=56$, $T=1$ triplet, and its mass excess can be calculated using 
\begin{equation}
\Delta M = a + bT_z + cT_z^{2}.
\end{equation}
The $a$ coefficient for integer triplets is the mass excess of the isobaric analogue state (IAS) of the $T_z$~=~0 member of the triplet, in this case the $J^{\pi}=4^+$ state in $^{56}$Ni, and can be calculated from the reported IAS excitation energy of 6432~keV \cite{Borcea200169}. The IMME $b$ and $c$ coefficients for the $A=56$ triplet have not been published, but can be estimated using fits to coefficients of triplets in the vicinity of $A=56$. Global fit functions of IMME parameters have been discussed in \cite{MacCormick201461},  where the authors treat the nucleus as a homogeneous charged sphere, and coefficients $a$, $b$ and $c$ are reported for the $A =4n$ subgroups. Here, we fit only to coefficients for a local region with A$=$32, 36, 40 and 48. As per the homogeneous charged sphere approximation of \cite{PhysRev.147.735}, the $b$ and $c$ coefficients can be parametrized in the following manner:
\begin{align}
b &= C^1_b - C^2_b \times (A - 1) / A^{-1/3} \\
c &= C^1_c + C^2_c \times A^{-1/3}
\end{align}
where $C^1_b, C^2_b, C^1_c, C^2_c$ are fit parameters. The fits obtained for $b$ and $c$ in the local vicinity are then used for the $A = 56$, $T=1$ subgroup. The resulting fit extrapolated to $A = 56$ results in $c$ = 110(95)~keV and $b$ = -8680(109)~keV.  Along with the result for the $a$ coefficient from \cite{Borcea200169} of 6431.9 (7) ~ keV, this provides a mass excess prediction for $^{56}$Cu of -38685(82)~keV and, thus, a Q-value of 639 $\pm$ 82~keV. The error is taken from the largest deviation between a measured mass and the predicted value from the fit function in the local region of interest.

\begin{table}[h]
\caption{Summary of predictions for the Q-value of $^{55}$Ni$(p,\gamma)$.}
\label{table2}
\begin{ruledtabular}
\begin{tabular}{ccp{2.4cm}}
Q-value (keV) & Method & Reference \\ \hline 
\\
560 $\pm$ 140 & Mass extrapolation & AME2003 \cite{Wapstra2003129}\\
190 $\pm$ 200 & Mass extrapolation & AME2012 \cite{doi:1880784} \\
600 $\pm$ 100 & Coulomb Shift / Shell Model & Brown et al. \cite{PhysRevC.65.045802} \\
639 $\pm$ 82 & IMME & This work 
\end{tabular}
\end{ruledtabular}

\end{table}

As seen in Table \ref{table2}, the more precise estimate from this work agrees within errors with the Coulomb-shift calculation from \cite{PhysRevC.65.045802}, favoring a higher Q-value compared to the lower extrapolated value reported in the AME2012 compilation. A recent IMME-based estimate using the T=2 quintet \cite{Tu201689} reported a Q-value of 651(88)~keV. Moreover, requiring reasonable Coulomb shifts for higher-lying mirror states, as extracted experimentally in \cite{Orrigo} between $^{56}$Cu and $^{56}$Co, also favors a higher Q-value.  

\section{Thermonuclear reaction rate}
With our measurement and our predicted $^{56}$Cu mass, we have determined the resonance energies of the $^{55}$Ni(p,$\gamma$)$^{56}$Cu reaction. In order to determine the astrophysical reaction rate, proton- and $\gamma$-widths ($\Gamma_p$ and $\Gamma_\gamma$ respectively) were calculated for each state using a shell-model with the GXPF1A interaction \cite{GXPF1A} (Table \ref{table1}). These calculations allowed up to 3-particle 3-hole excitations in the $pf$-shell.

\begin{table*}[th!]
\caption{New measured and shell-model excitation energies for $^{56}$Cu up to 3~MeV, resonance energies ($E_r$), and tentative spin-parity assignments. Spectroscopic factors $C^2S$ used to calculate the partial proton and gamma widths ($\Gamma_p$ and $\Gamma_\gamma$ respectively) were calculated utilizing a shell model calculation with the GXPF1A interaction, using experimental energies when available.}
\centering
\begin{ruledtabular}
\begin{tabular}{ccp{0cm}ccccccc}
 \multicolumn{2}{c}{Experiment} & &  \multicolumn{2}{c}{Shell Model} & & \multicolumn{2}{c}{$C^2S$} & & \\ 
$E_x$ (keV) & $E_{r}$ (keV) & & $E_x$ (keV) & $E_{r}$ (keV) & $J^\pi$& $l=1$ & $l=3$ & $\Gamma_\gamma$ (eV) & $\Gamma_p$ 
(eV) \\
166(1) & & & 146 & & $(3_1^+)$ & $ 0.84 $  &  $9.1 \times 10^{-3}$ & $ 8.4 \times 10^{-5}$ &  \\
572(1) & & & 483 & & $(5_1^+)$ & 0.70 & 0.16 & $1.1 \times 10^{-3}$ &  \\
826(3) & 187(82) & & 1066 & 427 & $(4_2^+)$ & 0.12 & 0.69 & $4.5 \times 10^{-4}$ & $1.2 \times 10^{-16}$ \\
1037(3)& 398(82) & & 1023 & 384 & $(2_1^+)$ & 0.64 & 0.16 & $1.2 \times 10^{-2}$ & $1.7 \times 10^{-7}$ \\
1224(4)& 585(82) & $\bigg \{$ &  \begin{tabular}{@{}c@{}} 1146 \\ 1474 \end{tabular} &  \begin{tabular}{@{}c@{}} 507 \\ 835 \end{tabular} & \begin{tabular}{@{}c@{}} $(5_2^+)$ \\ $(3_2^+)$ \end{tabular} &  \begin{tabular}{@{}c@{}} 0.15 \\ 0.10 \end{tabular} &  \begin{tabular}{@{}c@{}} 0.71 \\ 0.68 \end{tabular} &  \begin{tabular}{@{}c@{}} $2.0 \times 10^{-3}$ \\  $1.9 \times 10^{-3}$ \end{tabular} & \begin{tabular}{@{}c@{}} $4.3 \times 10^{-6}$ \\  $3.9 \times 10^{-5}$ \end{tabular} \\
& & & 1582 & 943 & $0^+_1$ &  & $3.8 \times 10^{-2}$ & $1.6 \times 10^{-6}$ & $1.5 \times 10^{-4}$ \\
& & & 1913 & 1274 & $2^+_2$ & $ 0.15 $ & 0.57 & $1.4 \times 10^{-2}$ & $3.7$ \\
& & & 2036 & 1397 & $1^+_1$ &  & $1.3 \times 10^{-2}$ & $4.5 \times 10^{-4}$ & $ 8.1 \times 10^{-3}$ \\
& & & 2066 & 1427 & $3^+_3$ & 0.59  & $9.1 \times 10^{-2}$ & $2.2 \times 10^{-2}$ & 48 \\
& & & 2226 & 1587 & $2^+_3$ & $1.7\times 10^{-3} $ & $3.9 \times 10^{-2}$ & $3.8 \times 10^{-3}$ & 0.53 \\
& & & 2272 & 1633& $4^+_3$ & 0.63 & 0.13 & $5.3 \times 10^{-2}$ & 210\\
& & & 2350 & 1711 & $7^+_1$ & & $9.9 \times 10^{-3}$ & $1.2 \times 10^{-4}$ & $5.8 \times 10^{-2}$ \\
& & & 2393 & 1754 & $6^+_1$ & & 0.72 & $3.1 \times 10^{-2}$ & 5.5 \\
& & & 2419 & 1780 & $1^+_2$ & & $1.0 \times 10^{-2}$ & $9.8 \times 10^{-3}$ & $8.8 \times 10^{-2}$ \\
& & & 2483 & 1844 & $3^+_4$ & $5.5\times 10^{-2} $& $1.3 \times 10^{-2}$ & $8.9 \times 10^{-3}$ & 59 \\
& & & 2505 & 1866 & $1^+_3$ &  & $7.3 \times 10^{-3}$ & $2.0 \times 10^{-2}$ & 0.11 \\
& & & 2543 & 1904 & $2^+_4$ & $1.6\times 10^{-2} $ & $7.5 \times 10^{-3}$ & $9.2 \times 10^{-3}$ & 23 \\
& & & 2630 & 1991 & $3^+_5$ & $8.8\times 10^{-3} $ & $3.6 \times 10^{-3}$ & $9.6 \times 10^{-3}$ & 19 \\
& & & 2723 & 2084 & $4^+_4$ & $2.3\times 10^{-2} $ & $2.0 \times 10^{-2}$ & $9.9 \times 10^{-3}$ & 75 \\
& & & 2762 & 2123 & $6^+_2$ &  & $5.0 \times 10^{-2}$ & $1.5 \times 10^{-2}$ &  2.6 \\
& & & 2914 & 2275 & $5^+_3$ & $1.1\times 10^{-2} $& $1.0 \times 10^{-2}$ & $1.7 \times 10^{-2}$ &  77 \\

\label{table1}
\end{tabular}
\end{ruledtabular}
\end{table*}

\begin{figure}[h]
\centering
\includegraphics[width = 0.50\textwidth]{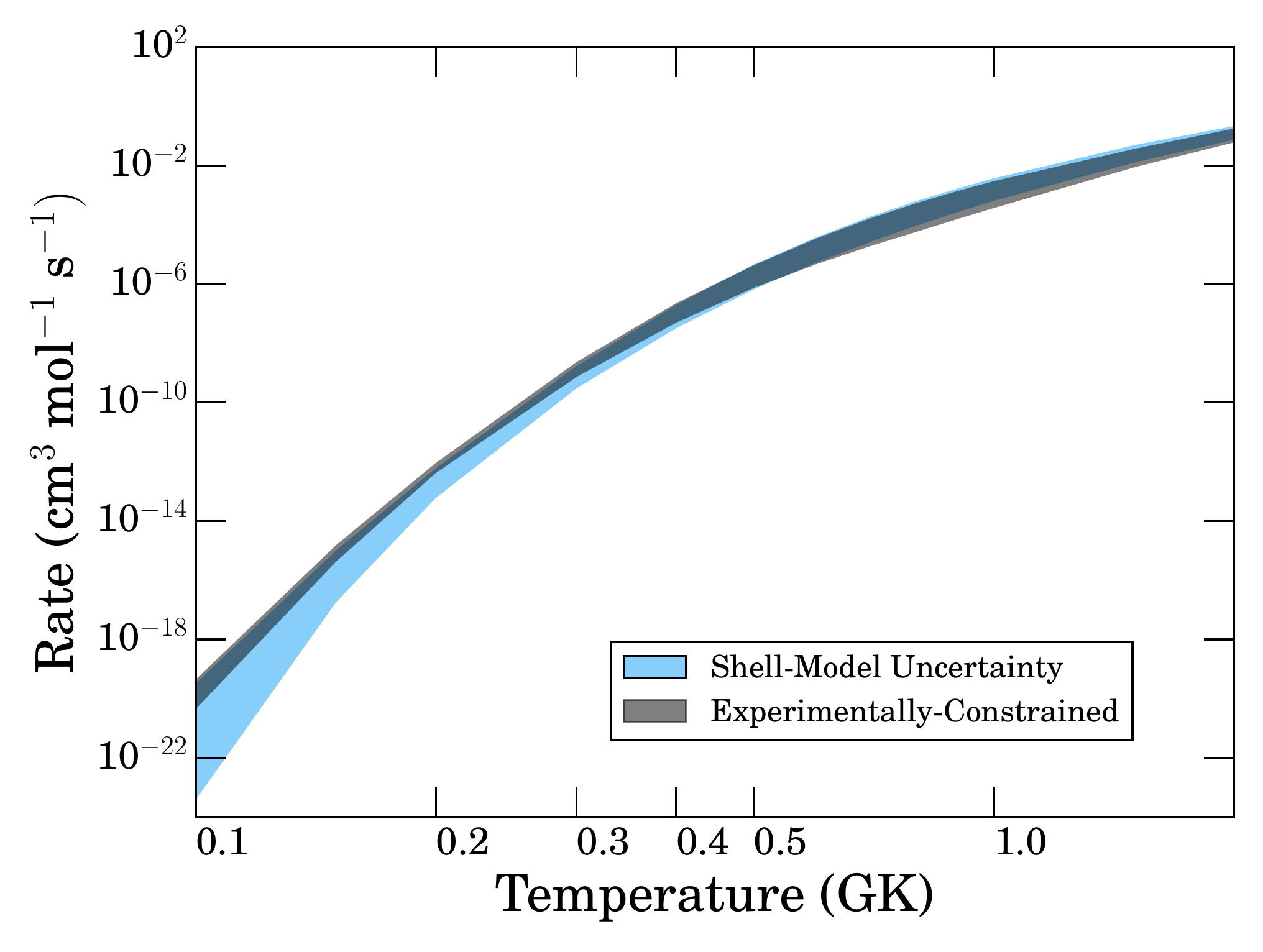}
\caption{(Color online:) Rate predictions showing the reduction of rate uncertainty by this work, assuming Q = 639~(82)~keV. We only consider uncertainties from resonance energy errors. The light band (blue) shows the 1-$\sigma$ uncertainty in the shell model rate, whereas the dark band (grey) shows the 1-$\sigma$ uncertainty in the experimentally-constrained rate. A clear reduction of the rate uncertainty in the temperature region of interest can be seen, especially at lower temperatures.} 
\label{Rate617}
\end{figure}

Reaction-rate uncertainties were calculated with a Monte-Carlo approach, similar to that of \cite{0954-3899-42-3-034007}, to properly account for the uncertainties in the excitation energies. Resonance energies and the reaction Q-value were allowed to vary assuming a Gaussian distribution within the uncertainties given in Table \ref{table1}. The uncertainty in the spin assignment for the 1224~keV state was also taken into account, but this represented only a small percentage of the uncertainty. The sampled resonance energy and corresponding rescaled proton-widths are used as input to Eq.~\ref{rateeq}, producing a sample of rates. At a given temperature, the 50$^{\mathrm{th}}$, 16$^{\mathrm{th}}$ and 84$^{\mathrm{th}}$ percentiles of the distribution of rate values provides the median, and 1-$\sigma$ uncertainty, respectively. The results are shown in Fig. \ref{Rate617}. To assess the reaction rate uncertainty prior to our measurement, we used the shell-model calculation and assumed a 200~keV uncertainty for the resonance energies. The resulting rate uncertainty (the light blue band in Fig. \ref{Rate617}) ranges from 4 orders of magnitude at 0.1~GK to about an order of magnitude at 2.0~GK. This is reduced at low temperatures to less than two orders of magnitude by our measurement (the gray band in Fig. \ref{Rate617}). The additional uncertainty from the calculated proton and $\gamma$ partial widths is estimated to be significantly smaller, about of a factor of 2  \cite{PhysRevLett.113.032502}. Thus, the dominant remaining source of uncertainty is the $\sim$80~keV error in the $^{56}$Cu mass, with smaller contributions from the uncertainties of the experimentally-unmeasured proton and $\gamma$ partial widths. 

Table \ref{reaclibtab} gives the corresponding REACLIB rate fit coefficients, using the parametrization given in Eqn. \ref{reaclibpara}, for our updated $^{55}$Ni(p,$\gamma$) reaction rate.

\begin{equation}
\begin{split}
N_A < \sigma \upsilon > &=  \sum_{i} {\mathrm {exp}}(a_{0i} + a_{1i}T_9^{-1} + a_{2i}T_9^{-1/3} + a_{3i}T_9^{1/3} \\
&\quad + a_{4i}T_9 + a_{5i}T_9^{5/3} + a_{6i}{\mathrm {ln}}T_9)
\label{reaclibpara}
\end{split}
\end{equation}

\begin{table}[h]
\caption{The recommended reaction rate $N_A \langle \sigma \upsilon \rangle$ as a function of temperature $T$ (GK) from this work, together with 1-$\sigma$ uncertainties (higher and lower).}
\begin{ruledtabular}
\begin{tabular}{cccc}
T$_9$ & \multicolumn{3}{c}{$N_A \langle \sigma \upsilon \rangle$ (cm$^3$/s/mole)} \\ 
& Recommended & Lower & Upper \\ \hline
0.1 & 1.497e-19 & 8.583e-20 & 2.422e-19 \\
0.2 & 8.661e-12 & 8.121e-12 & 1.082e-11 \\
0.3 & 1.666e-08 & 1.183e-08 & 2.084e-08 \\
0.4 & 1.244e-06 & 7.796e-07 & 1.746e-06 \\
0.5 & 1.859e-05 & 1.237e-05 & 3.606e-05 \\
0.6 & 1.156e-04 & 8.254e-05 & 2.911e-04 \\
0.7 & 4.677e-04 & 3.197e-04 & 1.357e-03 \\
0.8 & 1.529e-03 & 8.877e-04 & 4.423e-03 \\
0.9 & 3.880e-03 & 2.009e-03 & 1.149e-02 \\
1.0 & 8.951e-03 & 4.287e-03 & 2.551e-02 \\
1.5 & 1.583e-01 & 8.221e-02 & 3.294e-01 \\
2.0 & 9.620e-01 & 6.177e-01 & 1.655e+00 \\
\end{tabular}
\end{ruledtabular}
\end{table}

\begin{table*}[]
\caption{REACLIB fit coefficients for our recommended $^{55}$Ni(p,$\gamma$) reaction rate.}
\centering
\begin{ruledtabular}
\begin{tabular}{cccccccc}
$E_x$ & $a_0$ & $a_1$ & $a_2$ & $a_3$ & $a_4$ & $a_5$ & $a_6$ \\ \hline
1224 & 1.052854 &-6.805068 & 7.127737E-01 & -1.049583 & 5.849955E-02 & -3.234916E-03 & -9.787774E-01 \\
Other& -5.223069E+01 &-9.902812 & 1.336866E+02 & -7.623392E+01 & -8.335959E-01 & 2.019964E-01 & 6.914259E+01  \\
1038 &-5.177171 & -4.627019 &  -7.755680E-02 & 8.817104E-02 & -4.086783E-03 & 1.981643E-04 & -1.549327 \\                                  
826 & -2.601956E+01 & -2.170262 & 4.521332E-04 & -6.347735E-04 & 3.674535E-05 & -2.248394E-06& -1.499681 \\                                 

\label{reaclibtab}
\end{tabular}
\end{ruledtabular}
\end{table*}
\section{Consequences on the rp-process flow around $^{56}$Ni}
The astrophysical conditions that would lead the rp-process flow to bypass the $^{56}$Ni waiting point were investigated using a limited reaction network that includes the nuclides in Fig. \ref{bypassflow}. The network was seeded with $^{55}$Ni, where the rp-process enters the A~=~56 region. $^{56}$Ni was treated as a sink in the network calculation, with only flow into this nuclide being allowed. In this case, the ratio of the abundance of all other nuclei ($^{57}$Ni, $^{57,58}$Cu, and $^{58}$Zn) to the total abundance in the $N~=~28$ and $N~=~29$ chains is a measure of the fraction of the rp-process reaction flow that bypasses $^{56}$Ni, as it measures the amount of material trapped in neither $^{55}$Co nor $^{56}$Ni. The reaction network was run at constant temperature and proton density for 1~s, approximately 5 half-lives of $^{55}$Ni. A constant proton density was ensured by keeping the mass density constant, and by using a large proton-to-seed ratio of $\sim$400 such that the change in the proton abundance due to the comsumption of protons is negligible. 

\begin{figure}[h]
\includegraphics[width=0.5\textwidth]{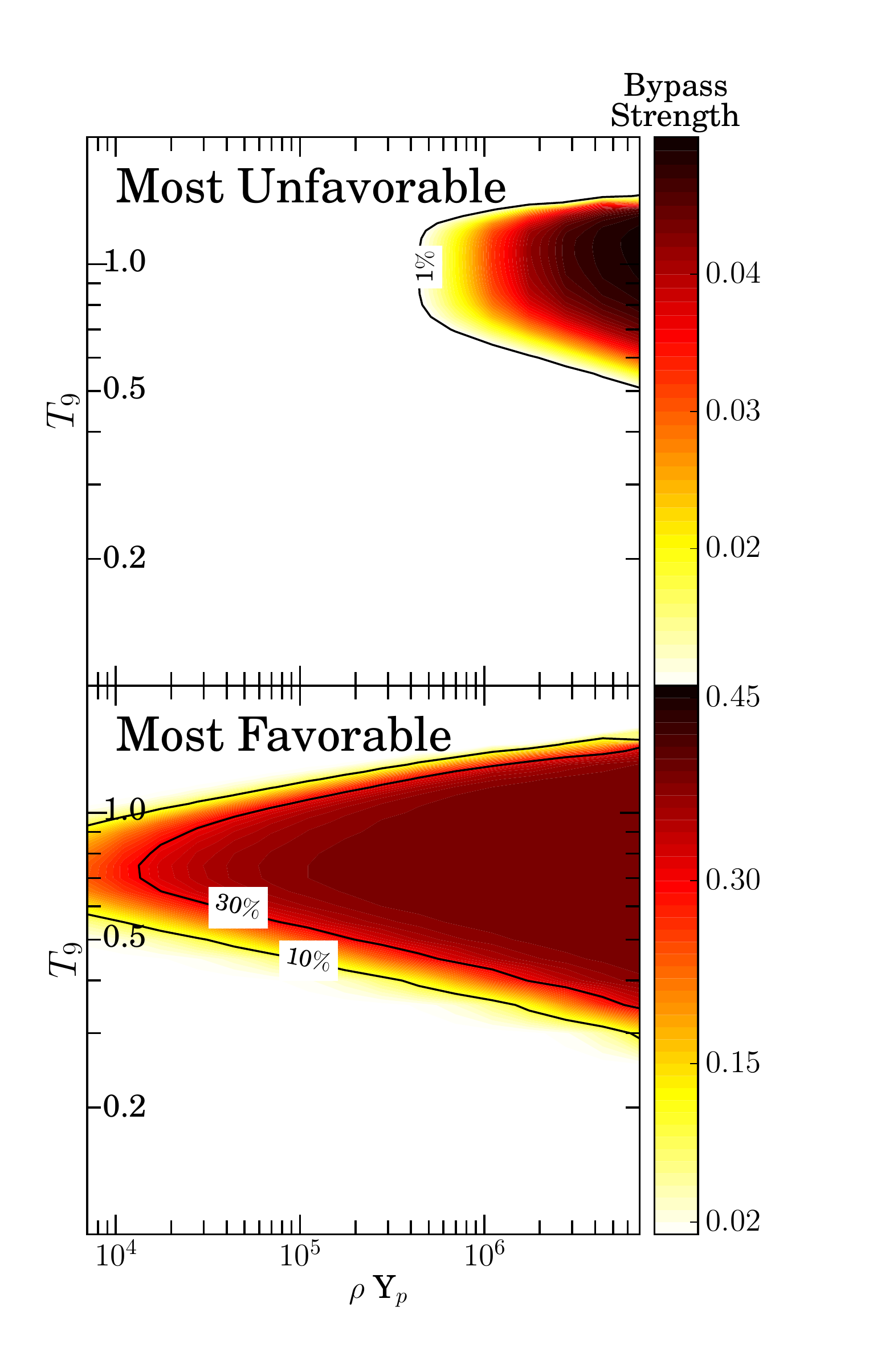}
\caption{(Color online:) Phase space diagram showing the region where the bypass may be effective, demonstrating the impact of the remaining nuclear physics uncertainties. The color and contours indicate the strength of the bypass. The most unfavorable (left) and most favorable (right) conditions are chosen to demonstrate the full range of the uncertainties.}
\label{ps}
\end{figure}
 
Even with the constraint on the $^{55}$Ni(p,$\gamma$) rate from this work, there remain additional uncertainties that affect the rp-process flow. The proton-capture rate on $^{56}$Cu determines the branching at $^{56}$Cu, where $\beta$ decay leads back to $^{56}$Ni, and also determines the total proton-capture flow at $^{55}$Ni in the case of $(p,\gamma)-(\gamma,p)$ equilibrium between $^{55}$Ni and $^{56}$Cu. In addition, the mass of $^{57}$Zn has not been measured and its uncertainty affects the $^{57}$Zn($\gamma$,p) rate, which hampers the flow bypassing $^{56}$Ni at high temperatures. Finally, the uncertain 78 $\pm$ 17 $\%$ $\beta$-delayed proton branch of $^{57}$Zn \cite{BBlank} directs the reaction flow back to $^{56}$Ni and needs to be better constrained. To explore the effect of these uncertainties, we considered two scenarios of maximal and minimal favorability for the bypass. In the case of the maximal (minimal) favorability: (1) the $^{56}$Cu(p,$\gamma$) rate was increased (decreased) by a factor of 100, the expected uncertainty of a shell-model rate; (2) the $^{55}$Ni(p,$\gamma$) rate was increased (decreased) by the uncertainty reported in this work; (3) the $\beta$-delayed proton-emission rate of $^{57}$Zn was decreased (increased) by the uncertainty reported by \cite{BBlank}. 
\\
Fig. \ref{ps} shows the resulting fraction of the reaction flow that bypasses $^{56}$Ni as a function of temperature and proton density for the two scenarios. In the scenario with the most favorable nuclear physics assumptions, $^{56}$Ni is significantly bypassed for temperatures in the range of about 0.4 - 1.2~GK and proton densities above 10$^4$~g/cm$^{3}$. These are within the range of typical X-ray burst conditions, with peak temperatures of 1-2~GK and proton densities up to 10$^6$~g/cm$^{3}$. On the other hand, for the most unfavorable scenario proton densities in excess of 10$^6$~g/cm$^{3}$ are required for the reaction flow to bypass $^{56}$Ni. Therefore, in the favorable scenario, $^{56}$Ni would be partially bypassed by the rp-process in all X-ray bursts, while in the unfavorable scenario the full rp-process would always pass through $^{56}$Ni. 

\section{Conclusion}

This work presents the first experimentally-constrained $^{55}$Ni(p,$\gamma$)$^{56}$Cu thermonuclear reaction rate, utilizing 5 newly identified excited states in $^{56}$Cu, a new theoretically-constrained reaction Q-value, and a new shell-model calculation of $\gamma$- and proton-widths. Below a temperature of 0.5~GK, the experimental data reduce the rate uncertainty from a factor of 10$^5$ to 10$^2$ at 0.1~GK and by almost an order of magnitude at 0.5~GK . The dominant remaining uncertainty is the reaction Q-value due to the unknown mass of $^{56}$Cu. For temperatures above 0.5~GK, the reaction rate is dominated by higher-lying resonances that have not been determined experimentally. With the new data, and using a detailed network analysis, we find that within remaining uncertainties the rp-process can bypass the $^{56}$Ni waiting point for typical x-ray burst conditions with a bypass branch as high as $\sim$40$\%$. We also identify additional nuclear physics uncertainties in the $^{56}$Cu(p,$\gamma$) reaction rate, the $^{57}$Zn mass, and the $^{57}$Zn $\beta$-delayed proton emission branch that need to be addressed. 
\\ \par
The authors want to thank the staff and the beam operators at the NSCL for their effort during the experiment. This work is supported by NSF Grants No. PHY11-02511, No. PHY10-68217, No. PHY14-04442, No. PHY08-22648 (Joint Institute for Nuclear Astrophysics), and No. PHY14-30152 (JINA Center for the Evolution of the Elements). GRETINA was funded by the U.S. DOE Office of Science. Operation of the array at NSCL is supported by NSF under Cooperative Agreement PHY11-02511 (NSCL) and DOE under Grant No. DE-AC02-05CH11231 (LBNL).

\bibliography{56cu}
\end{document}